# The Distances to Five Type II Supernovae Using the Expanding Photosphere Method, and the Value of $H_0$

Brian P. Schmidt and Robert P. Kirshner
Harvard-Smithsonian Center for Astrophysics, 60 Garden St., Cambridge, MA 02138

Ronald G. Eastman
Board of Studies in Astronomy and Astrophysics, Lick Observatory, University of California, Santa Cruz, CA 95064

Mark M. Phillips, Nicholas B. Suntzeff, Mario Hamuy
Cerro Tololo Inter-American Observatory[1], Casilla 603, La Serena, Chile

José Maza
Universidad de Chile, Dept. de Astronomia, Casilla 36-D, Santiago, Chile

Roberto Avilés
Cerro Tololo Inter-American Observatory[1], Casilla 603, La Serena, Chile

## Abstract

We have used observations gathered at CTIO to measure distances by the Expanding Photosphere Method (EPM) to 5 Type II supernovae. These supernovae lie at redshifts from $cz = 1100$ km s$^{-1}$ to $cz = 5500$ km s$^{-1}$, and increase to 18 the number of distances measured using EPM. We compare distances derived to 11 Type II supernovae with distances to their host galaxies measured using the Tully-Fisher method. We find that the Tully-Fisher distances average 11%±7% smaller. The comparison shows no significant evidence of any large distance-dependent bias in the Tully-Fisher distances. We employ the sample of EPM distances from 4.5 Mpc to 180 Mpc to derive a value for the Hubble Constant. We find $H_0 = 73 \pm 6$(statistical) $\pm$ 7(systematic) km s$^{-1}$ Mpc$^{-1}$.

[1]Cerro Tololo Inter-American Observatory, National Optical Observatories, operated by the Association of Universities for Research in Astronomy, Inc., (AURA) under cooperative agreement with the National Science Foundation.

## 1. Introduction

In recent years, work done on Planetary Nebula Luminosity Functions, Surface Brightness Fluctuations, and Line-Width Luminosity Relation has reached a consensus on the value of the Hubble constant, $H_0 = 80 \pm 10$ km s$^{-1}$ Mpc$^{-1}$ (Jacoby et al. 1992). Yet, good arguments for a "long" distance scale have been advanced by Sandage et al. (1992 and 1994) using SNe Ia as standard candles. On the basis of their Cepheid distance calibration of SN 1937C (IC 4182) and SN 1972E (NCG 5253) with HST, Sandage et al. (1994) obtain $H_0 = 56 \pm 8$ km s$^{-1}$ Mpc$^{-1}$. If $H_0$ really is 80 km s$^{-1}$ Mpc$^{-1}$, and $\Omega_0 = 1$ as preferred by many, then the 8 billion year expansion age of the Universe implied by $H_0$ ($\Lambda = 0$) is substantially younger than the oldest stars (Renzini 1991). Since there is disagreement among reliable workers, and our overall picture for cosmic evolution is at stake, it is useful to evaluate the distance scale using different approaches.

Kirshner and Kwan (1974), based on a suggestion by Leonard Searle, showed how SNe II could be used to measure extragalactic distances in a way completely independent of all other steps in the cosmic distance ladder. This technique, the Expanding Photosphere Method (EPM), was applied to two nearby SNe II and produced distances consistent with $H_0 = 60$ km s$^{-1}$ Mpc$^{-1}$. This work, which assumed SNe II emit like perfect blackbodies, was extended by Wagoner (1977), who demonstrated that EPM could, in principle, be applied to SNe II at large redshifts to measure $q_0$. Branch et al. (1983) and Kirshner (1985) applied EPM to additional SNe II, but the limitations of these primitive results of EPM were highlighted by Wagoner (1981), who argued that SNe II have scattering-dominated atmospheres, and therefore radiate with a smaller surface flux than a blackbody with the same color temperature. EPM was lifted to solid theoretical and empirical ground after several groups produced sophisticated NLTE models of SN 1987A's atmosphere (Eastman & Kirshner 1989; Höflich 1988; Schmutz et al. 1990; Chilkuri and Wagoner 1988). Application of EPM using these results gave distances to the LMC which agree with those derived using Cepheid variable stars (Walker 1987). Schmidt, Kirshner and Eastman (1992) [SKE92] exploited the tools developed for SN 1987A to measure the distances to ten SNe II, and found these distances again consistent with $H_0 = 60 \pm 10$ km s$^{-1}$ Mpc$^{-1}$. It is interesting to note that the exclusion of blackbody correction factors by Kirshner & Kwan (1974) did not significantly affect the derived distances to their galaxies. Because the correction factors change as a SN evolves, their inclusion not only alters the estimated surface flux of the SN, but also the time of explosion — two effects which can cancel in certain situations. Eastman, Schmidt and Kirshner (1994) [ESK94], have improved the basis for these distances by calculating a large grid of models which are not specific to SN 1987A and its blue supergiant progenitor. The relation between color temperature and flux inferred from these models generally produces distances 10% smaller than those based on the cruder formulation by SKE92. The distance measurements reported here use these improved models. Recently, Schmidt et al. (1994a) demonstrated the range over which EPM could be applied by developing a technique for applying EPM to SNe II at large redshifts, and measuring the distance to SN 1992am at a redshift of $cz = 14\ 600$ km s$^{-1}$.

We present new EPM distances to five SNe II observed at Cerro Tololo Inter-American Observatory (CTIO) as part of the Calan/Tololo SN search and monitoring program being carried out as a collaborative effort between the University of Chile and CTIO (Hamuy et al. 1993). Details of the observations will be published separately by Phillips et al. (1994), Schmidt et al. (1994b), and Hamuy et al. (1994). Redshifts for the host galaxies of these five SNe II range between 1100 km s$^{-1}$ and 5500 km s$^{-1}$. When combined with previous studies, our new data bring the total number of EPM distances to 18. These distances now extend from the LMC ($D$ = 49 kpc) to SN 1992am ($D$ = 180 Mpc) — an unparalleled range of distances for an extragalactic distance indicator. The host galaxies of these SNe II are spirals which, if not too distant or face-on, have distances measured using the Line-Width Luminosity Relation (Tully and Fisher 1977). We compare EPM and Tully-Fisher distances to 11 galaxies which have been measured both ways. We present our preliminary results on the value of $H_0$ from 16 SNe, and comment on the future role of EPM in determining the extragalactic distance scale.

## 2. Methods

EPM has been discussed in detail by SKE92. They show the photospheric angular size of a SNe II is given, for $z << 1$, by

$$\theta = \frac{R}{D} = \sqrt{\frac{f_\lambda}{\zeta_\lambda^2 \pi B_\lambda(T_\lambda)}}, \tag{1}$$

where $T$ is the SN's observed color temperature, $f_\lambda$ is the observed flux density, $B_\lambda(T)$ is the Planck function evaluated at $T$, and $\zeta_\lambda$ is a distance correction factor derived from model atmospheres to account for the dilution effects of scattering atmospheres. SNe II expand freely at velocity, $v$ (measured from the absorption minima of optically thin lines such as Fe II 5169), so that a supernova's photospheric radius, $R$, at any time, $t$, is

$$R = v(t - t_0) + R_0. \tag{2}$$

The initial radius,$R_0$, is negligible at all but the earliest epochs, and combining equations (1) and (2) yields

$$t = D\left(\frac{\theta}{v}\right) + t_0. \tag{3}$$

Given at least two measurements of $t$, $\theta$, and $v$, determined from a calendar, photometry, and spectra, it is possible to solve for both the distance to the SN, $D$, and the time of explosion, $t_0$, simultaneously. It is also possible to test the performance of the method and constrain $D$ through equation (3) if several measurements of $t$, $v$, and $\theta$ are available; despite large changes in $T$, $v$, $R$, and $\theta$, the distance should remain constant.

SKE92 demonstrated that the distance correction factors, $\zeta_\lambda$, as derived empirically and from a very limited set of models, varied only as a function of color temperature, and not from supernova to supernova. ESK94 investigated the effects of metallicity, progenitor mass, luminosity, density, and physical structure on the distance correction factor with a grid of 88 models, and found that $\zeta_\lambda(T)$ is not sensitive to details of the progenitor star. However, the new distance correction factors of ESK94 give 10% smaller distances on average than those determined from the models of SKE92. SKE92 relied strongly on the distance correction factors derived from SN 1987A's $VI$ photometry in their formulation of $\zeta_\lambda(T)$. These distance correction factors were applied to objects for which only $BV$ photometry was available. Although the values of $\zeta_\lambda(T)$ determined from $BV$ and $VI$ are similar, they are not identical, and this small difference is responsible for much of the systematic difference in the distances derived. In addition, SN 1987A, which resulted from the explosion of a compact progenitor, had a slightly different behavior of $\zeta_\lambda(T)$ than typical SNe II with larger initial radii. The custom-crafted models created for SN 1987A (Eastman and Kirshner 1989) are not quite in the center of the distribution of models, which cover a a wide range of plausible initial conditions, computed by ESK94. The distance correction factors determined from models by ESK94 are superior to those of SKE92, and we adopt their formulation of $\zeta_\lambda(T)$.

SNe II occur in dusty galaxies with recent star formation, and the effects of extinction cannot be ignored. SKE92 demonstrated that because a SN II's photosphere is formed in recombining hydrogen, most SNe II have a uniform color evolution during their photospheric phases ($\sim$ 100 days following explosion), and it is possible to estimate their color excess, $E(B-V)$, to an accuracy of 0.1 mag. Furthermore, the distances derived using EPM are affected by extinction in two conflicting ways (attenuation and reddening of the light) which cancel to a large extent. The uncertainty in a distance from extinction is usually less than 10%. In a worst case example, the distance to SN 1973R, which had $2.7 \pm 0.5$ magnitudes of visual extinction, has a distance uncertainty due to extinction of about 20% (SKE92). EPM distances are not uniformly increased or decreased by extinction, but instead depend on the evolution of the SN and the spacing of the observations. Errors in the reddening are unlikely to bias the average result even when they add noise to individual distances.

We estimate our errors using two techniques: Monte Carlo simulations and bootstrapping (Press *et al.* 1992). Our Monte Carlo simulations estimate the uncertainty of $D$ and $t_0$ from thousands of synthesized data sets which are constructed from the observable quantities, the modeled parameters, and uncertainty in these items. We assume the photometry has a gaussian error distribution of width equal to the quoted uncertainty, typically less than 0.05 mag. The gaussian distributed error in a velocity, characteristically around 300 km $s^{-1}$, is estimated from the scatter of velocities measured from different lines such as Fe II 5169, Fe II 5018, Sc II 5526, and Sc II 5658. The values of $\zeta_\lambda$ are selected at random from a uniform (not gaussian) distribution over the complete range of values determined in the models of ESK94. We believe this attributes a realistic uncertainty to the match of the model atmospheres to the true properties of the radiating stars.

As an alternative, bootstrapping does not require prior error estimates for any of the parameters, but instead it relies on the actual scatter of the data for error estimates. Because bootstrapping is essentially a Monte Carlo simulation where the actual data set has replaced the synthesized data, it does not work well in cases where there are few data points (in our case, simultaneous measurements of $T$, and $v$), and not at all when there are only two data points. We prefer using the bootstrap method, and apply it whenever there are more than five data points. The agreement between the two error estimates for cases with more than 6 data points is usually within 25%, which suggests we have reasonable estimates for the errors of the quantities in our Monte Carlo simulations.

The methods of error determination described here have the shortcoming that some of the uncertainty in EPM is related to systematic differences between the derived and actual values of parameters such as $v$, and $\zeta_\lambda$. Except in pathological cases, these errors will cause $\theta/v$ vs. $t$ to deviate from a straight line — if EPM is suffering severe systematic problems, they will be apparent. However, small systematic shifts, which are not time dependent, will not be so obvious, and require a more thoughtful analysis. Our best estimate, on the basis of the spread in distances independently derived to SN 1987A, is that EPM gives distances with systematic errors less than $\pm 10\%$.

Aside from the measuring accuracy of the photometry and spectra, the precision of EPM depends on the age of the SN at the time of the first observations and the spacing of subsequent observations. Observations which begin within three weeks of the explosion date, and which continue on a regular basis (once per week) afterwards, usually tightly constrain the time of explosion and significantly reduce the uncertainty in a derived distance. The statistical error in a distance determined from a set of 5 or more observations, spaced a week apart, and of good quality (e.g. 5% photometry, absorption features clearly visible), is smaller than the systematic uncertainties in EPM. In these cases, increasing the precision of the photometry and quality of the spectra, or increasing the frequency at which these measurements are obtained, will not improve the accuracy of the derived distances. Here we present results based on new observations of 5 supernovae, some of which have sufficiently good data that the estimated error in the distance approaches our uncertainty in the systematics of the method. While we only quote the measured error for each case, we renew our caution on the systematic effects when we use the results from all the objects to estimate $H_0$.

## 3. Distance Measurements

### a. SN 1986L

On 1986 Oct 7.6 the Rev. Robert Evans discovered SN 1986L in NGC 1559 (Evans 1986). Photometry taken soon after discovery showed SN 1986L was a very blue object still on the rise. The $B$ and $V$ light curves of this SN dropped off from the maximum rather quickly and then slowed their rate of descent producing a pronounced plateau. This characteristic is typical of many type II supernovae, and defines the SN II-P(lateau) subclass. The initial spectrum of SN 1986L was essentially featureless, and although it soon developed absorption lines, the spectrum never developed strong hydrogen P-Cygni features. SN 1986L was exceptionally well observed spectroscopically and photometrically ($B$ and $V$), and will be discussed in detail by Phillips *et al.* (1994). Given SN 1986L's initial blue color, (B-V) = -0.10, the low inclination of its host galaxy, and the small amount of reddening implied by the Galactic H I column depth in its direction (Burstein and Heiles 1984), we conclude this object had very little extinction. The spectrum of SN 1986L is peculiar, and not well represented by the grid of models calculated by ESK94. However, if we assume that this SN's distance correction factors behave like typical SNe II, we derive a distance of $16 \pm 2$ Mpc to the SN and its host galaxy, NGC 1559 (Table 1).

**b. SN 1989L**

SN 1989L was discovered by the Berkeley Automated Supernova Search on 1989 Jun 1.4 in the highly inclined galaxy NGC 7331. The object was visible on pre-discovery CCD frames taken by the Berkeley group as early as 1989 May 4 (Pennypacker and Perlmutter 1989), and remained at a nearly constant brightness until falling off the plateau in mid-September. SN 1989L shows classic P-Cygni lines throughout its early evolution, although the derived velocities (1700 km $s^{-1}$ at the end of the plateau) are smaller than for most other SNe II. A detailed discussion of this object as well as SN 1990K will be given in another paper (Schmidt *et al.* 1994b). Due to the high inclination of its host galaxy, it is not surprising SN 1989L appears to suffer from a significant amount of extinction as evidenced by its red color on the plateau, and a strong narrow Na D absorption line at the velocity of the host galaxy. SKE92 argue on theoretical and empirical grounds that most SNe II have a similar color evolution during their photospheric phases. From the (B-V) and (V-I) colors, relative to well observed SNe II such as SN 1969L (Ciatti, Bertola, and Rosino 1971) and SN 1987A (Hamuy et al. 1988), we estimate the visual extinction to SN 1989L to be $A_V = 1.0^m \pm 0.4^m$. The observations of SN 1989L, which are of high quality but have poor coverage, give a distance to NGC 7331 of $17 \pm 4$ Mpc using EPM (Table 2). Notice that, although the extinction to SN 1989L has a 40% uncertainty, the estimated error in the distance from all sources is only about 25%.

**c. SN 1990K**

The Rev. Robert Evans discovered SN 1990K as it emerged from the sun on 1990 May 25.8 (Evans 1990) in NGC 150. Our initial spectrum shows that this supernova was a typical SN II caught several weeks past maximum. SN 1990K's $B$,$V$,$R$, and $I$ light curves fall at a nearly a constant rate of 2.5 mag/100 days for the first 35 days after discovery, before falling sharply (~1.8 mag in 20 days) to the radioactive tail. This constant rate of decline is characteristic of the SN II-L(inear) subclass; however, the prototypes of this subclass, SN 1979C and SN 1980K, did not fall sharply to the radioactive tail, and this object may not fit neatly into either subclass. SN 1990K's host galaxy, NGC 150, lies near the south galactic cap, and the extinction from our Galaxy is minimal (Burstein and Heiles 1982). However, SN 1990K occurred in the middle of a spiral arm, and it is likely that there is some reddening from the host galaxy. We estimate the visual extinction to SN 1990K to be $A_V = 0.5^m \pm 0.4^m$ from the $(B - V)$ and $(V - I)$ color of the light curve compared to SN 1969L and SN 1987A. The observations of SN 1990K are extensive and of high quality, and we use them to derive a distance of $20 \pm 5$ Mpc to NGC 150 (Table 3). The principal uncertainty in this distance results from the age of the SN at the time of the first observations.

### d. SN 1992af

SN 1992af was discovered by R. Antezana as part of the Calan/Tololo supernova search on 1992 Jun 29 (Wells 1992). The first spectrum revealed hydrogen lines with classic P-Cygni profiles as well as weak Fe II lines indicating the SN was discovered about two weeks past maximum. $B$, $V$, and $I$ photometry show that SN 1992af remained on the plateau for 50 days following discovery, before falling approximately one magnitude to the radioactive tail. The SN's host galaxy, ESO 340-G38, is nearly face on, and should not be significantly obscured by dust in our Galaxy (Burstein and Heiles 1982); therefore we assume the visual extinction to SN 1992af is zero. The $B$, $V$, and $I$ light curves of SN 1992af are extremely well defined, but we were only able to obtain two spectra, the minimum to measure a distance. We derive a distance to the SN of $55^{+25}_{-20}$ Mpc (Table 4).

### e. SN 1992ba

On 1992 Sep 30.8 the Rev. Robert Evans discovered SN 1992ba (Evans 1992) in NGC 2082. The first spectrum obtained of this object shows well developed, strong P-Cygni profiles of the hydrogen Balmer series and He I 5876 indicating this was a SN II caught near maximum. The youth of the SN at discovery is further supported by a prediscovery $I$ plate, taken with the U.K. Schmidt Telescope on 1992 Sep 17, which shows no sign of the supernova to 19th magnitude (McNaught 1992). SN 1992ba's $B$, $V$, and $I$ light curves follow a typical pattern for SNe II, remaining relatively constant for 100 days following maximum, and then falling approximately 2 magnitudes to the radioactive tail. SN 1992ba's host galaxy, NGC 2082, lies at relatively low galactic latitude, $b = -33°$, and H I column depth measurements correspond to a visual extinction of $A_V = 0.11^m$ (Burstein and Heiles 1982). The SN occurred near (or in) a bright H II region in the galaxy, and may have additional extinction from the host galaxy. This is supported by Ca II H&K and Na D lines observed at the recession velocity of the galaxy which are slightly stronger than those at zero velocity. We estimate the visual extinction to SN 1992ba to be $A_V = 0.3^m \pm 0.2^m$. SN 1992ba was extremely well-observed both photometrically and spectroscopically and these observations give a distance to NGC 2082 of $14 \pm 1.5$ Mpc (Table 5).

## 4. Discussion

The five applications of EPM presented here bring the total number of SNe II with measured distances to 18 (Table 6). Some of the host galaxies for these SNe II are close enough that EPM can be compared to distances measured using conventional techniques such as Cepheids, while others are at twice the distance to Coma, so that perturbations in the Hubble flow should be small. In between, we can make a galaxy-by-galaxy comparison with Tully-Fisher distances. Three galaxies have Cepheid and EPM distances in common, the LMC (SN 1987A), M 81 (SN 1993J), and M 101 (SN 1970G). In the case of the LMC and M 101, the distances derived by the Cepheids, $49 \pm 4$ kpc (Walker 1987) and $7.1 \pm 0.3$ Mpc (Cook, Aaronson, and Illingworth 1986) respectively, are indistinguishable from those measured with EPM, $49 \pm 6$ kpc (Eastman and Kirshner 1989) and $7.4^{+1.0}_{-1.5}$ Mpc (ESK94). However, EPM appears to give a distance to M 81, $2.6 \pm 0.4$ Mpc (Schmidt et al. 1993), which is 25% smaller than that derived using Cepheids, $3.6 \pm 0.4$ Mpc (Freedman et al. 1993). It should be noted that SN 1993J has an atypical spectroscopic and photometric evolution which suggests the progenitor lost most of its hydrogen envelope before exploding. These observations, combined with evidence for asymmetry (Januzzi et al. 1993), suggest applying EPM to this object demands extreme care and models specific to the event. Baron, Hauschildt, and Branch (1993) have specifically modeled SN 1993J, and derive a distance to the SN in perfect agreement with the Cepheid distance. As several groups produce custom crafted models of SN 1993J, we will better understand these distance estimates to this unusual event. Two other SNe II, SN 1979C and SN 1986L, also have peculiar spectroscopic features (although not to the extent of SN 1993J), as noted in Table 6, and caution is clearly in order with respect to the distances to these two SNe II.

Table 6 shows eleven galaxies which have EPM and Tully-Fisher (TF) distances in common. Following Pierce (1993), we plot the distances determined using the two methods in Figure 1, and find good correlation. However, the TF distances of the sample average 11% ±7% smaller than those determined using EPM. This difference, which is marginally significant, is not surprising given that both Tully-Fisher and EPM have systematic uncertainties approaching 10% due to calibration (in the case of TF) and uncertainties in the physics of SNe II (EPM). More interestingly, the slope of Figure 1 is unity, within the estimated uncertainties, showing that there is no systematic difference between the distances derived by T-F and EPM nearby and far away. Because EPM does not make a standard candle assumption and is therefore immune to the classic Malmquist bias symptoms, this indicates that the sample of T-F distances here also has no significant distance-dependent error. This conclusion differs from the view presented by Sandage (1994), who attribute the large value of $H_0$ obtained through Line-Width Luminosity relations to Malmquist bias.

Observed large scale perturbations in the Hubble flow (Aaronson et al. 1982; Lynden-Bell et al. 1988; Mathewson et al. 1992; Lauer and Postman 1993) make measuring $H_0$ from nearby galaxies a risky proposition. The Virgo cluster significantly affects the local Hubble flow (Aaronson et al. 1982), and most of our EPM sample have distances comparable to or smaller than the distance to the Virgo Cluster. After correcting the observed heliocentric redshifts of each galaxy for Galactic rotation (220 km $s^{-1}$), we apply a non-linear infall model (Schechter 1980; SKE92) to the sample using a range of parameters discussed by Huchra (1988). As shown in Table 6, except for the two SNe which lie in the Virgo Cluster, SN 1979C and SN 1988A, the choice of these parameters does not strongly affect the derived recession velocities for most of the galaxies. SN 1987A and SN 1993J have redshifts which are near zero or negative, and we exclude these objects in our determination of $H_0$. Using the 16 EPM distances in Table 6, and assigning 300 km $s^{-1}$ errors (representing uncertainties due to large scale peculiar motions not associated with Virgo infall) to all recession velocities, we derive a value of $H_0 = 73{\pm}6$ (statistical) km s$^{-1}$ Mpc$^{-1}$ (Figure 2), using the least-squares technique for data sets with errors in both directions described by Press et al. (1992). The value of $H_0$ derived from our sample depends somewhat on the velocity errors assigned to each galaxy, ranging from $H_0 = 70 \pm 3$ km s$^{-1}$ Mpc$^{-1}$, if velocity errors are assumed to be zero, to $H_0 = 75 \pm 7$ km s$^{-1}$ Mpc$^{-1}$ if we assign errors of 600 km s$^{-1}$ to each velocity measurement. The nearby galaxies in our sample give a smaller value of $H_0$ than those further away, and their weight depends on the error adopted. It is difficult to assign a precise overall error to our value of $H_0$ because the statistical uncertainty of the measurement (8%) is of the same order as possible systematic uncertainties in EPM.

Systematic errors in EPM arise from uncertainties in our understanding of SNe II physics. All of the distances used in this determination of $H_0$ are based on the atmospheric code developed by Eastman (Eastman and Kirshner 1989; Eastman and Pinto 1993). It may be possible to evaluate the accuracy of EPM by comparing the output of other atmospheric codes which make slightly different physical assumptions in computing model atmospheres. A limited comparison can be made in the case of SN 1987A, a SN which several groups modeled (Eastman & Kirshner 1989; Höflich 1988; Schmutz et al. 1990; Chilkuri and Wagoner 1988), where the agreement is better than ±10%. Assigning an underlying uncertainity of ±10% to EPM, we find $H_0 = 73 \pm 6$(statistical) $\pm$ 7(systematic) km s$^{-1}$ Mpc$^{-1}$.

An additional uncertainty in our value of $H_0$ is the possibility than the local expansion rate of the Universe is significantly different than the global rate (Turner, Cen, and Ostriker 1992). Our sample includes three galaxies which lie beyond $cz = 5000$ km s$^{-1}$. These galaxies give values of $H_0$ approximately 10% larger than those which lie at $cz <$ 2000 km s$^{-1}$, although the difference is hardly significant. Lauer and Postman (1992) find the value of $H_0$ does not vary more by more than ±7% over the range from 3000 km s$^{-1}$ to 15000 km s$^{-1}$ using Brightest Cluster Galaxies, suggesting that difference between the local and global values of $H_0$ may not be significant on the scales we have sampled.

## 5. Conclusions

We have measured distances to five SNe II using EPM, SN 1986L, SN 1989L, SN 1990K, SN 1992af, and SN 1992ba. These SNe range in distance from 14 to 55 Mpc, and increase the total number of SNe II with distances measured using EPM to 18. We compare the distances derived to 11 galaxies with both EPM and Tully-Fisher distances. The agreement is very good, although Tully-Fisher, on average, gives distances which are 11%±7% smaller than EPM. There is no systematic difference between the distances derived using the two methods to galaxies nearby and far away, indicating distance dependent effects are not strongly affecting the T-F distances to the galaxies we have in common. We use the our sample of EPM distances to derive a value of $H_0 =$ $73 \pm 6$(statistical) $\pm$ 7(systematic) km s$^{-1}$ Mpc$^{-1}$.

The importance of SN searches to this program cannot be overstated. None of the five objects presented here was discovered accidentally; instead each was the result of a systematic search for supernovae. As these searches continue, they will enable us to apply EPM to many SNe II at redshifts larger than $cz = 5\,000$ km s$^{-1}$. These objects, which will typically be between $17 < m_v < 20$ mag, combined with those already measured, will allow us to derive a value of $H_0$ which is robust against large-scale perturbations in the Hubble flow. In addition, there is no technical reason why we cannot apply EPM to SNe II discovered at redshifts beyond $z = 0.1$, giving us the possibility of a direct approach to measuring the deceleration parameter, $q_0$.

M. Hamuy and J. Maza gratefully acknowledge financial support from grant 92/0312 from Fondo Nacional de Ciencias y Tecnologia (Fondecyt-Chile). Supernova research at the Harvard University is supported by NSF grant AST 92-18475 and NASA grants NAG 5-841, and NGT-51002.

**Table 1. Observed Quantities for SN 1986L**

| Julian Date (2446000+) | $T^a_{BV}$ (K) | $\theta^a_{BV}$ ($10^{15}$ cm Mpc$^{-1}$) | $v_{ph}$ (km s$^{-1}$) | $\zeta_{BV}$ |
|---|---|---|---|---|
| 714.3 | 14200 | 0.0278 | 8300 | 0.40 |
| 715.3 | 14700 | 0.0272 | 8200 | 0.40 |
| 715.8 | 16900 | 0.0236 | 8180 | 0.38 |
| 716.2 | 12200 | 0.0342 | 8140 | 0.38 |
| 716.7 | 15600 | 0.0248 | 8100 | 0.39 |
| 717.3 | 12400 | 0.0336 | 8055 | 0.38 |
| 718.2 | 13600 | 0.0302 | 7980 | 0.40 |
| 729.3 | 8600 | 0.0501 | 7080 | 0.40 |
| 730.4 | 7800 | 0.0591 | 6990 | 0.44 |
| 732.4 | 6800 | 0.0720 | 6830 | 0.54 |
| 735.2 | 7100 | 0.0640 | 6400 | 0.50 |
| 736.2 | 5800 | 0.0960 | 6400 | 0.75 |
| 737.4 | 6600 | 0.0763 | 6300 | 0.57 |
| 738.4 | 6100 | 0.0848 | 6200 | 0.68 |
| 740.3 | 5900 | 0.0868 | 6100 | 0.72 |
| 742.3 | 5600 | 0.0920 | 5950 | 0.81 |

**Table 2. Observed Quantities for SN 1989L**

| Julian Date (2440000+) | $T^a_{BVI}$ (K) | $\theta^a_{BVI}$ ($10^{15}$ cm Mpc$^{-1}$) | $v_{ph}$ (km s$^{-1}$) | $\zeta_{BVI}$ |
|---|---|---|---|---|
| 7684.9 | 8800 | 0.0409 | 4600 | 0.41 |
| 7685.9 | 8900 | 0.0399 | 4600 | 0.41 |
| 7700.8 | 7000 | 0.0545 | 3400 | 0.49 |
| 7701.8 | 6900 | 0.0554 | 3400 | 0.50 |
| 7784.7 | 4500 | 0.1018 | 1700 | 0.82 |

[a]Corrected for $A_V = 1.0^m$

**Table 3. Observed Quantities for SN 1990K**

| Julian Date (2440000+) | $T^a_{BVI}$ (K) | $\theta^a_{BVI}$ ($10^{15}$ cm $\mathrm{Mpc}^{-1}$) | $v_{ph}$ km $\mathrm{s}^{-1}$ | $\zeta_{BVI}$ |
|---|---|---|---|---|
| 8042.9 | 6100 | 0.0682 | 5100 | 0.60 |
| 8044.9 | 6000 | 0.0682 | 5000 | 0.61 |
| 8050.9 | 5600 | 0.0736 | 4650 | 0.66 |
| 8054.9 | 5500 | 0.0735 | 4320 | 0.67 |
| 8057.9 | 5500 | 0.0729 | 4080 | 0.68 |
| 8063.8 | 5200 | 0.0781 | 3600 | 0.72 |
| 8067.8 | 5000 | 0.0823 | 3330 | 0.75 |
| 8071.9 | 5000 | 0.0782 | 3100 | 0.75 |

[a]Corrected for $A_V = 0.5^m$

**Table 4. Observed Quantities for SN 1992af**

| Julian Date (2448000+) | $T_{BVI}$ (K) | $\theta_{BVI}$ ($10^{15}$ cm $\mathrm{Mpc}^{-1}$) | $v_{ph}$ (km $\mathrm{s}^{-1}$) | $\zeta_{BVI}$ |
|---|---|---|---|---|
| 813.0 | 6550 | 0.0178 | 7000 | 0.54 |
| 833.6 | 5620 | 0.0226 | 4300 | 0.66 |

**Table 5. Observed Quantities for SN 1992ba**

| Julian Date (2440000+) | $T^a_{BVI}$ (K) | $\theta^a_{BVI}$ ($10^{15}$ cm $\mathrm{Mpc}^{-1}$) | $v_{ph}$ km $\mathrm{s}^{-1}$ | $\zeta_{BVI}$ |
|---|---|---|---|---|
| 8896.8 | 12300 | 0.0447 | 6700 | 0.44 |
| 8904.8 | 9600 | 0.0659 | 5000 | 0.41 |
| 8905.8 | 9500 | 0.0666 | 4800 | 0.41 |
| 8908.8 | 8500 | 0.0770 | 4400 | 0.41 |
| 8922.8 | 6300 | 0.0860 | 3400 | 0.58 |
| 8940.8 | 5400 | 0.0947 | 2700 | 0.69 |
| 8941.8 | 5400 | 0.0941 | 2700 | 0.69 |
| 8956.8 | 5100 | 0.0737 | 2300 | 0.74 |

[a]Corrected for $A_V = 0.3^m$

**Table 6. Distances to SNe II**

| SN | Galaxy | $v_{rec}^{a}$ (km $s^{-1}$) | $v_{rec}^{b}$ (km $s^{-1}$) | $A_V$ (mag) | T-F Distance[c] (Mpc) | EPM Distance (Mpc) |
|---|---|---|---|---|---|---|
| SN 1968L | NGC 5236 | 380 | 380 | 0.11 | $4.8^{+1.0}_{-0.8}$ | $4.5^{+0.7}_{-0.8}$ |
| SN 1969L | NGC 1058 | 620 | 620 | 0.18 | $9.2^{+1.3}_{-1.3}$ | $10.6^{+1.9}_{-1.1}$ |
| SN 1970G | NGC 5457 | 380 | 390 | 0.44 | $6.9^{+1.8}_{-1.4}$ | $7.4^{+1.0}_{-1.5}$ |
| SN 1973R | NGC 3627 | 450 | 460 | 2.70 | $7.6^{+1.1}_{-1.0}$ | $15^{+7}_{-7}$ |
| SN 1979C[d] | NGC 4321 | 1300 | 1150 | 0.45 | $14.5^{+2.9}_{-2.5}$ | $15^{+4}_{-4}$ |
| SN 1980K | NGC 6946 | 320 | 310 | 1.20 | $5.5^{+1.1}_{-0.9}$ | $5.7^{+0.7}_{-0.7}$ |
| SN 1986L[d] | NGC 1559 | 1110 | 1110 | 0.00 | $15.2^{+2.2}_{-2.0}$ | $16^{+2}_{-2}$ |
| SN 1988A | NGC 4579 | 1300 | 1150 | 0.00 | $16.6^{+2.5}_{-2.1}$ | $20^{+3}_{-3}$ |
| SN 1989L | NGC 7331 | 1420 | 1440 | 1.00 | $20.5^{+3.0}_{-2.6}$ | $17^{+4}_{-4}$ |
| SN 1990E | NGC 1035 | 1230 | 1240 | 1.70 | $13.7^{+2.1}_{-1.7}$ | $18^{+3}_{-2}$ |
| SN 1990K | NGC 150 | 1420 | 1440 | 0.50 | $13.7^{+2.1}_{-1.7}$ | $20^{+5}_{-5}$ |
| SN 1990ae | anon0020+06 | 7800 | 7800 | 0.50 | ... | $115^{+35}_{-25}$ |
| SN 1992H | NGC 5377 | 2240 | 2150 | 0.00 | ... | $31^{+4}_{-4}$ |
| SN 1992af | ESO 340-G38 | 5380 | 5400 | 0.00 | ... | $55^{+25}_{-20}$ |
| SN 1992am | anon0122-04 | 14500 | 14500 | 0.30 | ... | $180^{+35}_{-25}$ |
| SN 1992ba | NGC 2082 | 950 | 940 | 0.30 | ... | $14^{+1.5}_{-1.5}$ |

[a]Corrected for Virgo Infall; $V_{Vir}$ = 1050 km $s^{-1}$, $V_{in}$ = 250 km $s^{-1}$
[b]Corrected for Virgo Infall; $V_{Vir}$ = 950 km $s^{-1}$, $V_{in}$ = 200 km $s^{-1}$
[c]From Pierce (1993)
[d]Has Peculiar Spectroscopic Features

## References

Aaronson, M. et al. 1982, ApJS, 50, 241.

Baron, E., Hauschildt, P. H., & Branch, D. 1993 ApJL submitted.

Branch, D., Falk, S. W., McCall, M. L., Rybski, P., Uomoto, A. K., & Wills, B. J.1983., ApJ, 244, 780.

Burstein, D. & Heiles, C. 1982, AJ, 87, 1165.

Burstein, D. & Heiles, C. 1984, ApJS, 54, 33.

Chilkuri, M. & Wagoner, R.V. 1988, in Atmospheric Diagnostics of Stellar Evolution, IAU Colloquium 108, ed. K. Nomoto, (Berlin: Springer-Verlag) p295.

Ciatti, F., Rosino, L., & Bertola, F. 1971, MSAIt, 42, 163.

Cook, K. H., Aaronson, M., & Illingworth, G. 1986, ApJ, 301, L45.

Eastman, R. G. & Kirshner, R. P. 1989, ApJ, 347, 771.

Eastman, R. G., Schmidt, B. P., & Kirshner, R. P. 1994, To be submitted to ApJ

Eastman, R.G. & Pinto, P. A. 1993 ApJ, in press

Evans, R. 1986 I.A.U. Circular #4260.

Evans, R. 1990 I.A.U. Circular #5022

Evans, R. 1990 I.A.U. Circular #5625

Freedman, W. L. et al. 1993 Submitted to ApJ

Höflich, P. 1988, in Atmospheric Diagnostics of Stellar Evolution, IAU Colloquium 108, ed. K. Nomoto, (Berlin: Spinger-Verlag) p288.

Hamuy, M., Suntzeff, N. B., Gonzalez, R., & Martin, G. 1988 AJ, 95, 63.

Hamuy, M. et al. 1993 AJ in press

Hamuy, M. et al. 1994 in preparation

Huchra, J.P. 1988, in Proceedings of the A.S.P.: The Extragalactic Distance Scale, ed. S. van den Bergh & C.J. Pritchet, (Provo: Brigham Young University Press) p 257.

Jacoby, G. H., Branch, D., Ciardullo, R., Harris, W. E., Pierce, M. J., Pritchet, C. J., Tonry, J. L., & Welch, D. L. 1992, PASP., 104, 599.

Januzzi, B., Schmidt, G., Elston, R. & Smith, P. 1993 IAU Circular #5776

Kirshner, R. P. 1985, in Lecture Notes in Physics: Supernovae as Distance Indicators, ed. N. Bartel (Berlin: Springer-Verlag), p. 171.

Kirshner, R. P. & Kwan, J. 1974, ApJ, 193, 27.

Lauer, T. L. & Postman, M. 1992, ApJ, 400, L47.

Lauer, T. R. & Postman, M. 1993, in Texas/PASCOS 92: Relativistic Astrophysics and Particle Cosmology ed. C. W. Akerlof & M. A. Srednicki (Ann. N.Y. Acad. Sci) 688, 531.

Lynden-Bell, D., Faber, S. M., Burstein, D., Davies, R. L., Dressler, A., Terlevich, R. J., & Wegner, G. 1988, ApJ, 326, 19.

Mathewson, D. S., Ford, V. L., & Buchhorn, M. 1992, ApJ 389, L5

McNaught, R. H. 1992 I.A.U. Circular #5632

Pennypacker, C. & Perlmutter, S. 1989 I.A.U. Circular #4791

Phillips, M. M. et al. 1994 in preparation

Pierce, M. J. 1993 ApJ in press

Press, W. H., Flannery, B. P., Teukolsky, S. A., & Vettering, W. T. 1992. Numerical Recipes in C, 2nd ed. (Cambridge University Press: Cambridge)

Renzini, A. 1991. in Observational Tests of Cosmological Inflation, ed. T. Shanks et al. (Dordecht: Kluwer), p 313.

Sandage, A. 1994, ApJ submitted

Sandage, A. & Tammann, G. A. 1990 ApJ, 365, 1.

Sandage, A., Saha, A., Tammann, G. A., Panagia, N., & Machetto, F. D. 1992, ApJ, 401, L7.

Sandage, A., Saha, A., Tammann, G. A., Labhardt, L., Schwengeler, H., Panagia, N., & Machetto, F. D. 1994, ApJL submitted.

Schechter, P. L. 1980, AJ, 85, 801.

Schmidt, B. P., Kirshner, R. P., & Eastman, R. G. 1992, ApJ, 395, 366.

Schmidt, B. P., Kirshner, R. P., & Eastman, R. G., Grashuis, R., Dell'Antonio, I., Caldwell, N., Foltz, C., Huchra, J., & Milone, A. A. E., 1993, Nature, 364, 600

Schmidt, B. P., Kirshner, R. P., Eastman, R. G., Phillips, M. M., Suntzeff, N. B., Hamuy, M., Aviles, R., Filippenko, A. V., Ho, L., Matheson, T., Grashuis, R., Maza, J., Kirkpatrick, J. D., Kuijken, K., Zucker, D., Bolte, M., & Tyson, N. 1994a, AJ submitted

Schmidt, B. P. et al. 1994b in preparation

Schmutz, W., Abbot, D. C., Russell, R. S., Hamann, W. R., & Wessolowski, U. 1990, ApJ, 355, 255.

Tully, R. B. & Fisher J. R. 1977 A&A, 54, 661

Turner, E., Cen, R., & Ostriker, J. P. 1992, AJ, 103, 1427

Wagoner, R. V. 1977, ApJ, 214, L7

Wagoner, R. V. 1981, ApJ, 250, L65

Walker, A. R. 1987, MNRAS, 225, 627

Wells, L. 1992 I.A.U. Circular #5554

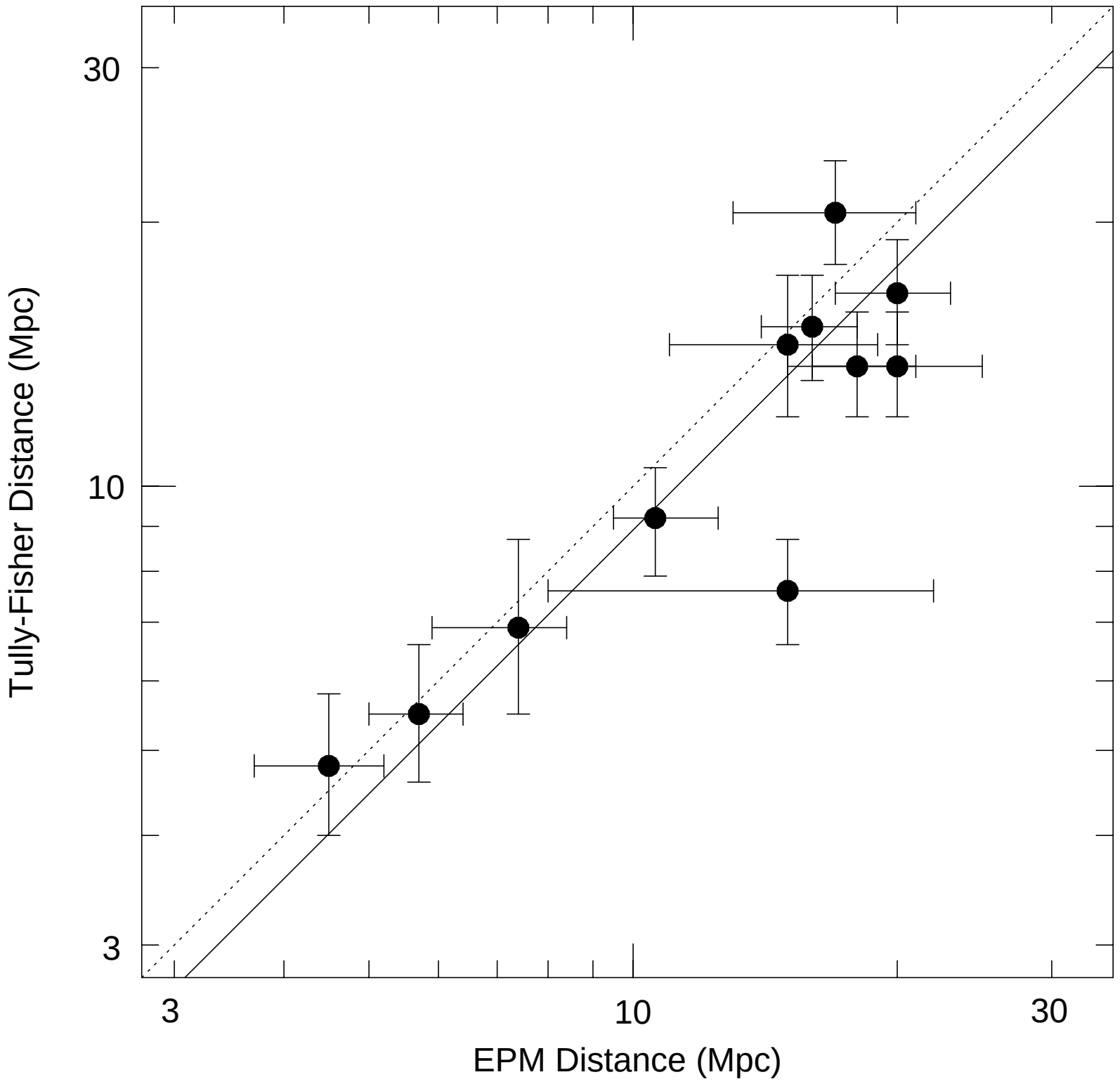

Figure 1: EPM distance versus Tully-Fisher distance is plotted for 11 galaxies. The solid line shows the best fit between the two samples, $D_{T-F} = 0.89 D_{EPM}$. The dashed line line shows $D_{T-F} = D_{EPM}$.

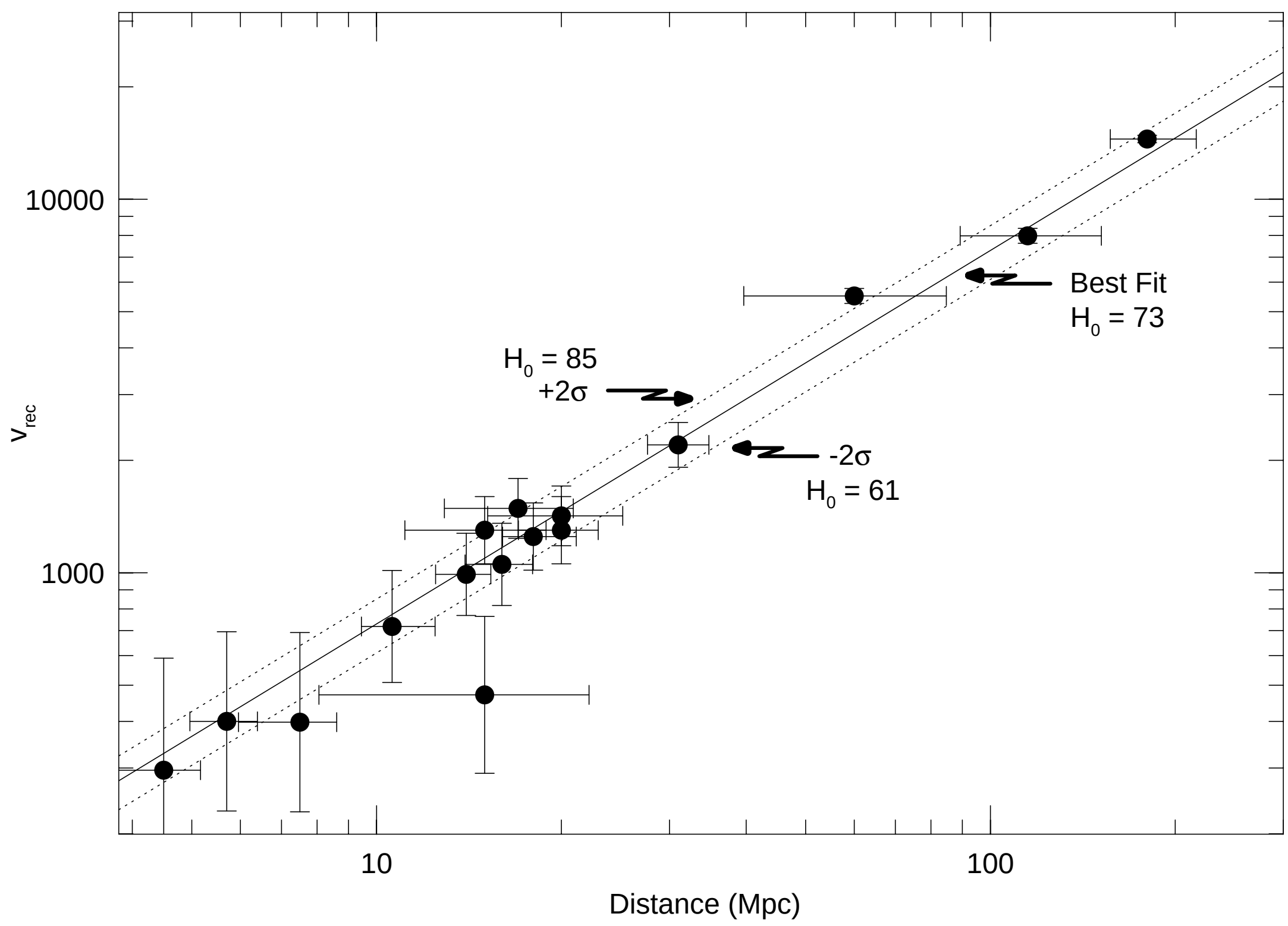

Figure 2: The EPM distances to 16 SNe II versus redshift (corrected for Virgo infall). Error bars give the 95% confidence limits for each distance. The solid line shows the best fit for the sample, $H_0 = 73$ km s$^{-1}$ Mpc$^{-1}$. The dashed lines give 2-$\sigma$ above and 2-$\sigma$ below the mean value.